\documentclass[a4paper,english]{article}
\usepackage{fullpage}
\usepackage{graphicx}
\usepackage{amsmath,amssymb}
\newcommand{\keywords}[1]{\textbf{Keywords:} #1}

\usepackage[affil-it]{authblk}  

\usepackage{hyperref}
\usepackage{cleveref}
\usepackage{xcolor}

\usepackage[nocompress]{cite} 
\usepackage{booktabs} 
\usepackage{graphicx} 
\usepackage{marvosym} 
\usepackage[algoruled, linesnumbered, noend]{algorithm2e}
\crefname{algocf}{alg.}{algs.}
\Crefname{algocf}{Algorithm}{Algorithms}

\newcommand{\eps}{\varepsilon}
\newcommand{\set}[1]{\left\{{#1}\right\}}

\title{Fast Public Transit Routing with Unrestricted Walking through Hub Labeling}

\author[1]{Duc-Minh Phan}
\author[2]{Laurent Viennot}

\affil[1]{Got It Vietnam, Hanoi, Vietnam\footnote{ORCID 0000-0003-4199-3101}\footnote{This work was mostly performed while this author was at Irif -- Paris University, France.}}
\affil[2]{Inria, Paris University, France\footnote{Supported by Irif CNRS laboratory and ANR project MULTIMOD (ANR-17-CE22-0016).}}

\begin{document}

\maketitle

\begin{abstract}
We propose a novel technique for answering routing queries in public transportation networks that allows unrestricted walking. We consider several types of queries: earliest arrival time, Pareto-optimal journeys regarding arrival time, number of transfers and walking time, and profile, i.e. finding all Pareto-optimal journeys regarding travel time and arrival time in a given time interval. Our techniques uses hub labeling to represent unlimited foot transfers and can be adapted to both classical algorithms RAPTOR and CSA. We obtain significant speedup compared to the state-of-the-art approach based on contraction hierarchies.

\end{abstract}

\keywords{Route planning, Public transportation, Hub labeling.}

\section{Introduction}
\label{s:intro}

Despite remarkable progress of route planning algorithms in road networks~\cite{road-survey}, public transit routing still requires specific algorithms due to its temporal nature. Various efficient methods were proposed such as CSA~\cite{csa}, RAPTOR~\cite{raptor}, Transfer Pattern~\cite{transfer-pattern,scalable-transfer-pattern}, PTL~\cite{ptl}. They all consider a graph with two types of edges: the connections that correspond to a vehicle traveling from a stop to the next one, and the transfers that correspond to walking from a stop to another nearby stop. While each connection is scanned only once per query, transfer edges from a stop are considered each time an event is detected at the stop. Efficiency of such techniques thus relies on the sparsity of the transfer graph. Additionally, they all share the requirement that the graph resulting from walking transfers is transitively closed and are generally experimented with a sparse transfer graph by restricting transfers to very short distances only. Allowing unrestricted transfers, that is walking from a stop to any other stop, is indeed out of reach with these methods although it would allow to find better answers. Indeed, recent work~\cite{DBLP:conf/atmos/WagnerZ17} shows the benefit of using unrestricted walking over sparse transfers by measuring that it can reduce travel time by hours in Switzerland and Germany networks.

This paper is devoted to enable unrestricted walking in efficient public transit routing. The motivation for considering unrestricted walking goes beyond the gain of quality in the answers. It is indeed a fundamental step towards computing multimodal journeys as it is considered as a main bottleneck in~\cite{multimod-raptor}. Note that bicycle or taxi transfers can be handled similarly as walking transfers with different speed and cost. Techniques developed for unrestricted walking can thus generalize to other modes of transportation.

A first step towards unrestricted walking was made by MCR~\cite{multimod-raptor} and UCCH~\cite{ucch} algorithms that both use a contracted version of the full walking graph, inspired by contraction hierarchies~\cite{CH}, for representing the full walking graph which is much bigger. However, this contracted graph is not transitively closed and has to be globally scanned several times during a query. Accelerating such computations with unrestricted walking is still challenging as multi-criteria or profile queries require seconds to be performed on practical networks with these methods.

In static graphs such as road networks, hub labeling~\cite{hub-labeling,hub-labeling-journal} (also called 2-hop labeling~\cite{cohen}) is a remarkable technique that achieves state-of-art response-time to shortest path queries. It consists in selecting for each node a small set of access nodes called hubs such that any shortest path can be described as a two hop travel through a common hub of the extremities. Intersecting the two lists of hubs of a source and a destination indeed allows to find efficiently the shortest path between them. Such technique was used in PTL~\cite{ptl} on the time expanded graph representation of a network to obtain fast transit routing. A similar approach is followed by TTL~\cite{ttl} which revisits hierarchical hub labeling~\cite{hierarchical-hub-labeling} in the context of public transit networks. However, these approaches still assume sparse transfers. Note that the time expanded graph representation duplicates transfer edges from a stop for all events at that stop, and its size can blow up with dense transfer graphs.

In this work, we propose a new approach for handling unrestricted walking in public transit routing based on a different usage of hub labeling. It basically consists in decomposing walking transfers into two consecutive hops. We use hub labeling in the classical setting of a static graph but in a novel manner compared to distance or shortest path queries: we scan hub lists to propagate reachability information. Interestingly, the technique can easily be adapted to both RAPTOR and CSA based algorithms which are the two main classical approaches with restricted transfers. HLRaptor, our variant of RAPTOR obtains significant speedup compared to MCR. HLCSA, our variant of CSA obtains competitive running times for earliest arrival time and profile queries.

The paper is organized as follows. Section~\ref{s:preliminaries} defines public transit networks, describes briefly RAPTOR and CSA algorithms, and introduces hub labeling. Sections~\ref{s:hlraptor} and~\ref{s:hlcsa} present HLRaptor and HLCSA respectively. We describe in Section~\ref{s:transit-data} public transit data used to evaluate our algorithms. The results of our experiments are presented in Section~\ref{s:experiments}.

\section{Preliminaries}
\label{s:preliminaries}

We define a public transit network with a triple $(S,T,R)$ representing trips of vehicles (buses, trains, etc.): $S$ is the set of \emph{stops} where passengers can enter or disembark from a vehicle, $T$ is the set of trips made by vehicles and are grouped into routes represented by the set $R$. More precisely, a \emph{trip} $t$ is given by a sequence of stops served by a vehicle and for each stop $u$ in the sequence, an arrival time $\tau_{arr}(t,u)$ of the vehicle at stop $u$ and a departure time $\tau_{dep}(t,u)$ of the vehicle from stop $u$. A \emph{route} $r$ consists in a set of trips with same stop sequence. This set of trips can be represented by a two-dimensional timetable where each line lists the arrival and departure times $\tau_{arr}(t,u),\tau_{dep}(t,u)$ of a trip $t$ for all stops $u$ in the sequence. Note that the sequence of times listed in a line is non-decreasing. Similarly to RAPTOR authors~\cite{raptor}, we assume that no trip of a route can overtake another trip of the same route. In other words, the lines of the timetable can be sorted so that each column is non-decreasing. This property can easily be enforced by splitting the set of trips with same stop sequence into smaller subsets of trips if necessary.

The public transit network is complemented by a weighted footpath graph $G=(V,E,\tau_w)$ with $S\subseteq V$, and $\tau_w(u,v)$ denotes the time needed to walk from a node $u$ to a node $v$. In the unrestricted walking setting, the graph is not assumed to be transitively closed and we expect that $V$ is much larger than $S$. Each edge $(u,v)\in E$ typically corresponds to a segment of street than can be traversed by walking. We let $d_G(u,v)$ denote the length of a shortest path in $G$ from $u$ to $v$, i.e., the minimum walking time from $u$ to $v$.

We consider several journey computation problems. Given a source stop $s$ and a target stop $t$, a \emph{journey} from $s$ to $t$ is an alternating sequence of trips and footpaths in the public transit network, which starts with $s$ and ends with $t$. The goal of public transit planning is to compute journeys from $s$ to $t$ optimizing one or several criteria. Given a departure time $\tau$, an \emph{earliest arrival time} query consists in computing a journey with minimum arrival time at $t$ that departs from $s$ at $\tau$ or later. In a \emph{multi-criteria} query, we are additionally interested in the number of transfers and the overall walking time of the journey, and ask for all Pareto-optimal journeys. Recall that a journey is Pareto-optimal if no other journey is better on one criterion and at least as good on all criteria. In a \emph{profile} (or \emph{range}) query, we ask for journeys whose departure time falls within a given time interval while optimizing both departure time and arrival time (where later departure time is considered better). The required answer again consists in all Pareto-optimal journeys within the given time interval.

The RAPTOR algorithm and its variants~\cite{raptor,multimod-raptor} compute journeys starting from a given stop at a given time in rounds, where each round extends partial journeys by one trip. More precisely, each round consists of two phases: the first phase explores each route of the public transit network and extends partial journeys arriving at stops served by a route using the first trip arriving at each stop. In the second phase, each partial journey arriving at a stop is extended by walking paths from that stop. In the regular RAPTOR version~\cite{raptor}, single-edge paths only are considered and the footpath graph is assumed to be transitively closed. In the unrestricted walking setting~\cite{multimod-raptor}, a multi-source Dijkstra is performed on a contracted version of the footpath graph in order to find all stops whose arrival times can be improved by walking from the stops that were scanned during the first phase.

The Connection Scan Algorithm (CSA)~\cite{csa} breaks each trip into consecutive \emph{connections}, which represent a vehicle traveling from a stop to the next one in the stop sequence of the trip. All connections are sorted by departure times in a pre-computation step. The algorithm scans all the connections and transfers to update the earliest arrival time at each reachable stops. More precisely, for each connection $c$ in increasing order of departure time, we need to check whether a passenger can travel on $c$ or not: either the trip containing $c$ has been reached earlier, or we can arrive at the departure stop of $c$ before its departure time. Then we update the arrival time at the arrival stop of $c$ if necessary, and scan the footpath transfers from the arrival stop of $c$. Similarly to RAPTOR, CSA also requires the footpath graph to be transitively closed.

\emph{Two-hop labeling}~\cite{cohen}, or equivalently, \emph{hub labeling}~\cite{hub-labeling,hub-labeling-journal}, for a (weighted, directed) graph $G$ consists in assigning two subsets of nodes $H_-(u)$ and $H_+(v)$ to each node $u$. Nodes in $H_-(u)$ (resp. $H_+(u)$) are called \emph{in-hubs} (resp. \emph{out-hubs}) and serve as intermediate nodes to reach $u$ (resp. to leave $u$). The following \emph{two-hop} property is required: for any pair $u,v$ of nodes, there must exist a common hub $h\in H_+(u)\cap H_-(v)$ lying on a shortest path from $u$ to $v$, i.e., satisfying $d_G(u,h)+d_G(h,v)=d_G(u,v)$. Equivalently, $H_+$ (resp. $H_-$) can be seen as a graph with vertex set $V$ and edges $(u,v)$ with weight $d_G(u,v)$ for every pair $u,v$ such that $v\in H_+(u)$ (resp. $u\in H_-(v)$). The two-hop property can then be stated as $H_+ \cdot H_- = G^*$, where $G^*$ denotes the transitive closure of $G$ and $\cdot$ denotes the graph product resulting from the $(\min,+)$-matrix product of adjacency matrices (the weight of an edge $(u,w)$ in $H_+ \cdot H_-$ is $\min_{v\in H_+(u)\cap H_-(w)} d_G(u,v)+d_G(v,w)$). In other words, any shortest path in $G$ corresponds to a two-hop path in $H_+\cup H_-$. The interest for such representation comes from the fact that it is possible to compute very small hub sets (less than 100 nodes on average) in large road networks and footpath graphs~\cite{massive}, and thereby obtain the fastest known practical oracles for computing distances and shortest paths in such networks~\cite{road-survey}.

\section{HLRaptor: RAPTOR with Two-Hop Transfers}
\label{s:hlraptor}

Using a hub labeling $H_-,H_+$ of the footpath graph $G$, we propose the following modification of RAPTOR that we call HLRaptor. We replace the second phase of a round by two sub-phases: in the first sub-phase we scan every stop $u$ for which arrival time $\tau_u$ was improved in the regular first phase of the round, and update arrival time at its out-hubs $h\in H_+(u)$ to $\min\set{\tau_h, \tau_u + \tau_w(u,h)}$. In the second sub-phase, we scan every hub $h$ whose arrival time was improved in the first sub-phase and update arrival time at nodes $v$ such that $h\in H_-(v)$ to $\min\set{\tau_v, \tau_h + \tau_w(h,v)}$.

The correctness of HLRaptor comes from the two-hop property of the hub labeling that ensures $H_+\cdot H_-=G^*$. Our two sub-phases using $H_+$ and $H_-$ are thus equivalent to the second phase of the regular RAPTOR algorithm using the transitive closure $G^*$ of $G$. However, its performance depends on the out-degrees of $H_+$ and $H_-$ rather than that of $G^*$.

\subsubsection{Target pruning optimization.} 
The lists $H_-^{-1}(h)=\set{v \mid h\in H_-(v)}$ and $H_+(u)$ can be pre-computed for all $u,h\in V$. Additionally, these lists can be sorted according to walking time from $u$ (resp. $h$) in non-decreasing order. This enables a target pruning optimization where we stop scanning a list as soon as the arrival time computed for a node in the list exceeds the best arrival time known at the target.

\subsubsection{HLprRaptor: profile queries with HLRaptor.}
We can follow the same approach as~\cite{DBLP:conf/atmos/WagnerZ17} to compute all Pareto-optimal journeys with respect to departure time and arrival time in a given interval of time. The difference is that we use HLRaptor instead of MCR. The idea is to use HLRaptor to compute the best arrival time $\tau_a$ when starting at a givent time $\tau$. Then we use a reverse version of HLRaptor (or simply a reversed version of the transit data) to compute the last departure time $\tau_d$ such that arrival at $\tau_a$ is still possible. We then repeat this procedure for departure time $\tau_d+\eps$ for sufficiently small $\eps$ (we simply use $\eps = 1$ second, which is the time unit in our datasets). We iterate this until all Pareto-optimal journeys in the given time interval have been found.

\subsubsection{HLmcRaptor: HLRaptor with multiple criteria.}
To deal with more criteria than arrival time and number of transfers, we can keep multiple non-dominating labels for each stop $u$ in round $k$ in a bag structure similarly to McRAPTOR~\cite{raptor}. For each route $r$ with a stop improved in the previous round, we scan the first trip departing after any improved arrival time at a stop $u$ of the route and update bags accordingly at the stops served by the trip after $u$. In the second phase of the round, each newly inserted label is first propagated along out-hubs links and then newly inserted labels at hubs are propagated along in-hubs links similarly. We can adapt local and target pruning as in McRAPTOR. We can also adapt our target pruning optimization specific to HLRaptor to stop scanning hub lists as soon as the propagated label is dominated by the destination bag.


\section{HLCSA: Connection Scan with Two-Hop Transfers}
\label{s:hlcsa}

Given a hub labeling $H_+,H_-$ of the walking graph $G$, we propose the following modification of CSA. For an earliest arrival time query from $s$ to $t$, we first scan out-hubs $H_+(s)$ and update arrival time to them by walking from $s$. Similarly to CSA, we then scan connections by non-decreasing departure time.
When considering a connection $c$, we first scan the in-hubs $H_-(u)$ of its departure stop $u$ and update the arrival time at $u$ through walking from a hub. The connection can be boarded if the trip has been marked as boarded or if the arrival time at $u$ plus the minimum transfer time at $u$ is no later than the departure time of $c$. In that case, we update the arrival time at the arrival stop $v$ of $c$ and scan its out-hubs $H_+(v)$ to update their arrival times through walking from $v$. Finally, we scan the in-hubs $H_-(t)$ of the destination $t$ and update the arrival time at $t$ by walking from any of them.

The correctness of the algorithm comes again from the two-hop property of hubs. For any possible transfer from a connection $c$ to another connection $d$ in a journey, $c$ must be considered before $d$. Let $h$ denote a common hub for the arrival stop $u$ of $c$ and the departure stop $v$ of $d$ such that $d_G(u,v)=d_G(u,h)+d_G(h,v)$ according to the two-hop property. After $c$ is considered, arrival time at $h$ is thus no more than $\tau+d_G(u,h)$, where $\tau$ is the arrival time of $c$ at $u$ and $d_G(u,h)$ is the walking time from $u$ to $h$. When $d$ is then considered, arrival time to $v$ is updated to $\tau+d_G(u,h)+d_G(h,v)=\tau+d_G(u,v)$ as if a transfer from $u$ to $v$ had been considered. A similar reasoning applies for a journey starting with a walk from $s$ or ending with a walk to $t$. HLCSA thus behaves as in a regular CSA execution where all transitive transfers in $G^*$ would be considered.

\subsubsection{Optimization.} In addition to all CSA classical optimizations, we can again sort out-hub lists in non-decreasing order of walking time, and apply target pruning similarly as in HLRaptor. In addition, we scan the in-hub list of the departure stop of a connection when the trip is not marked as boarded.
Again, this list can be sorted by non-decreasing walking time and we stop scanning the list as soon as the walking time from the hub exceeds the estimated travel time to the departure stop (local pruning).

\subsubsection{HLprCSA: Profile queries with HLCSA.}
Similarly to the original extension of CSA to solve the profile problem~\cite{csa}, we store for each stop a bag containing Pareto-optimal pairs of departure time at stop with associated arrival time at destination. We also store such information for hubs. We also consider connections in non-increasing order of departure time. When scanning a connection $c$, we use the bags of the out-hubs of its arrival stop to obtain the best arrival time through walking after $c$. If the arrival time of the trip of $c$ is improved, we then update the bags of the in-hubs of the departure stop of $c$ for that arrival time with departure time corresponding to walking from the hub for boarding right in time the connection. We also scan in-hubs of the destination at the beginning of the procedure and out-hubs of the source at end in order to take care of walking from source and to destination. The correctness of the modification follows similar lines as for HLCSA.

\section{Public Transit Data}
\label{s:transit-data}

To evaluate the algorithms, we use datasets from three locations: London, Paris, and Switzerland. The dataset for London was obtained from Transport for London~\cite{TFL}. The dataset for Paris was obtained from Open Data RATP~\cite{RATP}. And the dataset for Switzerland was provided by Wagner \& Z\"{u}ndorf~\cite{DBLP:conf/atmos/WagnerZ17}\footnote{\url{https://i11www.iti.kit.edu/PublicTransitData/Switzerland/}}.
The extracted dates are 2015-11-06 for London and 2018-03-30 for Paris.

The public transit data of Paris already has transfers between stops, we simply need to make the transfer graph transitively closed for appropriate use with RAPTOR and CSA. However, the dataset of London does not have transfers, thus we have created transfers by linking any pair of stops separated by 75 meters of walk one from another. This threshold was chosen to obtain a transitively closed transfer graph with similar size as in previous works. The graph obtained by transitive closure of restricted transfers is called \emph{transfer graph} in the sequel.

The footpath graphs for London and Paris were extracted from Geofabrik's data~\cite{Geo}, which is itself extracted from OpenStreetMap's data~\cite{OSM}. We call \emph{walking graph} the union of this unrestricted footpath graph and the transfer graph. The method to merge a stop of the public transit network into the walking graph is the following. For each stop $p$, we find the closest node $v$ in the walking graph. If the distance between $p$ and $v$ is less than 5 meters, we identify $p$ and $v$, connecting $p$ with the in- and out-neighbors of $v$ using the same weights. Otherwise, we find the 5 closest nodes of $p$ in the walking graph, and connect $p$ with those at distance 100 meters at most. If there are no nodes in the walking graph within the radius of 100 meters from $p$, then $p$ is isolated. Walking times are computed according to a walking speed of 4~km/h.   \Cref{tab:stats} provides  statistics concerning the datasets.
The columns stops and transfers provide the number of nodes and edges in the transitively closed restricted walking graph, while the last two columns give the numbers of vertices and edges in the unrestricted walking graphs.

\begin{table}[ht]
\def\arraystretch{1.2}
\setlength{\tabcolsep}{3pt}
\centering
\caption{Dataset statistics}
\label{tab:stats}
\begin{tabular}{lrrrrrrr}
            & routes & trips  & events  & stops & transfers & vertices & edges   \\
\hline
London      & 1622   & 122593 & 4695285 & 19746 & 46566    & 281167   & 840880  \\
Paris       & 1973   & 78757  & 1915253 & 23519 & 362291    & 533470   & 1666386 \\
Switzerland & 13930  & 369744 & 4740869 & 25427 & 38265     & 604230   & 1882551 \\
\end{tabular}
\end{table}

We computed hub labelings of the walking graphs using the sampling-based algorithm by Delling et al.~\cite{massive} (1-2h of pre-computation per graph).
\Cref{tab:trans} provides statistics on the degrees of transfer graphs vs. in-hubs and out-hubs graphs: $\delta^+(Tr)$ and $\Delta^+(Tr)$ designate the average and maximum out-degree resp. of the transfer graph $Tr$, $\delta^+(H_+)$ and $\Delta^+(H_+)$ designate the average and maximum out-degree resp. of the out-hub graph $H_+$, $\delta^-(H_-)$ and $\Delta^-(H_-)$ designate the average and maximum in-degree resp. of the in-hub graph $H_-$. We let $|H_+|$ and $|H_-|$ designate the number of edges in $H_+$ and $H_-$ resp. while $|V(H)|$ designates the number of hubs (including stops). We note that the size of hub lists is comparable to the number of events and their storage do not increase too much space requirements.

\begin{table}[ht]
\def\arraystretch{0.9}
\setlength{\tabcolsep}{2pt}
\centering
\caption{Transfers, out-hubs and in-hubs degrees.}
\label{tab:trans}
\begin{tabular}{lrrrrrrrrr}
            & $\scriptstyle\delta^+(Tr)$ & $\scriptstyle\Delta^+(Tr)$ & $\scriptstyle\delta^+(H_+)$ & $\scriptstyle\Delta^+(H_+)$ & $\scriptstyle\delta^-(H_-)$ & $\scriptstyle\Delta^-(H_-)$ & $\scriptstyle|V(H)|$ & $\scriptstyle|H_+|$ & $\scriptstyle|H_-|$ \\
  \hline
London      & 2.36        & 20         & 70       & 150   & 71       & 142
  & 65059 &  1393759 & 1395024  \\
Paris       & 15.4        & 205        & 118      & 196   & 118      & 210
  & 60519 & 2770336 & 2798315  \\
Switzerland & 1.5         & 26         & 78       & 229   & 79       & 230
  & 117793 & 2005312 & 2005312\\
\end{tabular}
\end{table}

We also prepared two sets of roughly 1000 queries for each dataset. In the first one, source and destinations are selected independently uniformly at random among all stops similarly to experiments in~\cite{raptor,csa,multimod-raptor}. In the second one, we select sources and destinations similarly to~\cite{DBLP:conf/atmos/WagnerZ17}: one hundred sources are selected uniformly at random. For each source, we order the destinations by increasing walking distance and select a random one uniformly among those with rank in $[2^i,2^{i+1}]$ for $i=2\ldots 14$. For Switzerland, we use exactly the same pairs as in~\cite{DBLP:conf/atmos/WagnerZ17} where sources are selected with probability proportional the number of trips serving them. In both sets, we additionally selected uniformly at random a departure time in $[0, 24 \times 3600]$ for each source-destination pair.
We will reference the two sets of queries as ``uniform'' and ``rank'' respectively. Note that most of the uniform queries (those in the uniform set) correspond to high rank pairs ($2^{12}$ or higher) while the rank set of queries has a strong bias towards low rank pairs.

The datasets are made publicly available\footnote{\url{https://files.inria.fr/gang/graphs/public_transport/}}.

\section{Experiments}
\label{s:experiments}

Our algorithms were implemented in C++ and compiled with GCC version 7.2.0 (with flag \texttt{-O3}). Experiments were conducted on one core of a dual 10-core Intel Xeon E5-2670-v2 with with 25~MiB of L3 cache and 64~GiB of DDR3-1866 RAM. The code is made available\footnote{\url{https://github.com/lviennot/hl-csa-raptor}}.

\begin{table}[ht]
\centering
\caption{Average running times of HLRaptor and HLCSA.
}
\begin{tabular}{l c@{}c@{}c@{}c @{\hskip 10pt} rrrr @{\hskip 10pt} rrrr}
\toprule
 &&&&& \multicolumn{4}{c}{London} 
     & \multicolumn{4}{c}{Paris} 
     \\
\cmidrule(lr){6-9} \cmidrule(lr){10-13} 
 &&&&& \multicolumn{2}{c}{Restricted} & \multicolumn{2}{c}{Unrestr.}
 & \multicolumn{2}{c}{Restricted} & \multicolumn{2}{c}{Unrestr.}
 \\
Algorithm & \smash{\rotatebox{70}{Range}\hskip -4pt} 
          & \smash{\rotatebox{70}{Arr.}} 
          & \smash{\rotatebox{70}{Nb. tr.}} 
          & \smash{\hskip -2pt\rotatebox{70}{Walk}} 
          & Unif. & Rank & Unif. & Rank
          & Unif. & Rank & Unif. & Rank      
          \\
\hline
HLRaptor & $\circ$ & $\bullet$ & $\bullet$ & $\circ$
  & 5.1 & 1.9 & 26.4 & 8.7
  & 3.0 & 1.3 & 19.5 & 5.5\\
HLCSA & $\circ$ & $\bullet$ & $\circ$ & $\circ$
  & 2.2 & 1.1 & 33.1 & 16.8
  & 1.0 & 0.5 & 13.8 & 6.5\\
HLmcRaptor & $\circ$ & $\bullet$ & $\bullet$ & $\bullet$
  & 87.3 & 33.0 & 417 & 140
  & 60.0 & 25.9 & 248 & 85.4\\
HLprRaptor (2h) & $\bullet$ & $\bullet$ & $\bullet$ & $\circ$
  & 76.5 & 31.1 & 685 & 237
  & 53.0 & 23.1 & 652 & 205\\
HLprCSA (2h)& $\bullet$ & $\bullet$ & $\circ$ & $\circ$
  & 47.1 & 28.4 & 1012 & 539
  & 60.9 & 35.4 & 628 & 330\\
HLprRaptor (24h) & $\bullet$ & $\bullet$ & $\bullet$ & $\circ$
  & 805 & 322 & 7522 & 2524
  & 567 & 262 & 7441 & 2511\\
HLprCSA (24h) & $\bullet$ & $\bullet$ & $\circ$ & $\circ$
  & 312 & 217 & 11644 & 8453
  & 404 & 298 & 9523 & 7902\\
\bottomrule
\end{tabular}
\label{tab:runtime}

\medskip

\begin{tabular}{l c@{}c@{}c@{}c @{\hskip 10pt} rrrr}
 &&&& 
     & \multicolumn{4}{c}{Switzerland} \\
\cmidrule(lr){6-9} 
 &&&&& \multicolumn{2}{c}{Restricted} & \multicolumn{2}{c}{Unrestr.}
 \\
Algorithm & \smash{\rotatebox{70}{Range}\hskip -4pt} 
          & \smash{\rotatebox{70}{Arr.}} 
          & \smash{\rotatebox{70}{Nb. tr.}} 
          & \smash{\hskip -2pt\rotatebox{70}{Walk}} 
          & Unif. & Rank & Unif. & Rank
          \\
\hline
HLRaptor & $\circ$ & $\bullet$ & $\bullet$ & $\circ$
  & 13.4 & 4.0 & 59.4 & 7.6\\
HLCSA & $\circ$ & $\bullet$ & $\circ$ & $\circ$
  & 6.6 & 2.9 & 54.2 & 19.4\\
HLmcRaptor & $\circ$ & $\bullet$ & $\bullet$ & $\bullet$
  & 150 & 62 & 854 & 229\\
HLprRaptor (2h) & $\bullet$ & $\bullet$ & $\bullet$ & $\circ$
  & 47.7 & 16.4 & 402 & 83.9\\
HLprCSA (2h)& $\bullet$ & $\bullet$ & $\circ$ & $\circ$
  & 51.9 & 32.1 & 563 & 240\\
HLprRaptor (24h) & $\bullet$ & $\bullet$ & $\bullet$ & $\circ$
  & 293 & 111 & 3461 & 751\\
HLprCSA (24h) & $\bullet$ & $\bullet$ & $\circ$ & $\circ$
  & 128 & 96 & 4173 & 3076\\
\bottomrule
\end{tabular}
\end{table}

\Cref{tab:runtime} presents the average running times in milliseconds of HLRaptor and HLCSA variants on the three datasets. We indicate for each algorithm which criteria are optimized: arrival time (Arr.), number of transfers (Nb. tr.), overall walking time (Walk), and whether the query spans a range of departure times (Range).

In the restricted walking setting, our algorithms are equivalent to the corresponding Raptor or CSA based version. On the London instance with restricted walking and uniform queries, we obtain similar results as Raptor~\cite{raptor} for earliest arrival, multi-criteria and 2h range queries: 5.1ms vs. 7.3ms, 87.3ms vs. 107ms, and 76.5ms vs. 87ms, respectively (we compare times reported in \Cref{tab:runtime} to times reported in \cite{raptor}). Our running times are 15-30\% faster, probably due to the use of more recent hardware. We also obtain similar results as CSA~\cite{csa} for earliest arrival and 24h range queries: 2.2ms vs. 1.2ms and 312 vs. 107ms. Our running times are 2-3 times slower than those reported in~\cite{csa}, probably due to less optimized code.

In the unrestricted walk setting, our algorithms are significantly faster than previous works. On the London instance with uniform queries and unrestricted walking, HLmcRaptor is 3.4 times faster than times reported for MCR in~\cite{multimod-raptor} (417ms vs. 1438ms) and HLRaptor is 1.7 times faster than the MR-$\infty$ variant of MCR (26.4ms vs. 44.4ms). On the Switzerland instance with ranked based queries and unrestricted walking, HLprRaptor computes profile queries roughly 7 times faster than the profile variant of MCR proposed in~\cite{DBLP:conf/atmos/WagnerZ17}: 751ms vs. 5.5s approximately. Most uniform queries have high rank, and HLprRaptor obtains their profiles in roughly 3.5s compared to 20s approximately as reported in~\cite{DBLP:conf/atmos/WagnerZ17}.

Interestingly, our hub-labeling-based versions of CSA obtain rather good performances with respect to Raptor based versions in the unrestricted walk setting: they are nearly as fast or even faster  on uniform queries, and at most 2-3 times slower on rank queries. (Note that on low rank queries, Raptor-based solutions benefit from target pruning.)

\begin{table}[ht]
\setlength{\tabcolsep}{7pt}
\centering
\caption{Average/median gain of unrestricted walking on travel time compared to restricted walking.}
\label{tab:gain}
\begin{tabular}{l cccc}
            & Unif. & Unif. 6h-20h & Rank & Rank 6h-20h\\
\hline
London      & 12\% / 5.8\% &  6.9\% / 2.9\% &  24\% / 13\% &  16\% / 5.0\%\\
Paris       & 22\% / 15\%  &  15\% / 13\%   &  31\% / 21\% &  22\% / 17\%\\
Switzerland & 47\% / 46\%  &  37\% / 39\%   &  47\% / 47\% &  35\% / 37\%\\
\end{tabular}
\end{table}

\subsubsection{Gain of unrestricted walking.}
We confirm the results of~\cite{DBLP:conf/atmos/WagnerZ17} showing the benefit of considering unrestricted walking.  \Cref{tab:gain} presents the percentage of time gained by using a journey with unrestricted walking compared to the travel time with restricted walking. The average gain ranges from 12\% to 47\% on uniform queries depending on the dataset. City networks (especially London) seem to benefit less from unrestricted walking than the train network of Switzerland. As observed in~\cite{DBLP:conf/atmos/WagnerZ17}, the gain is less important during daytime that is queries with departure time in the range 6h-20h here. We observe a higher gain on low rank queries. The median gain ranges from 13\% to 47\% for them. More precisely, the gain is at least 13\% on half of the low rank queries for London, 21\% for Paris and 47\% for Switzerland.

\section{Conclusion}

We have demonstrated the efficiency of using a two-hop representation of unrestricted walk transfers in conjunction with CSA and RAPTOR algorithm. This shows that is possible to enable unrestricted walking in practical public transit routing engines and opens new perspectives for allowing complex multimodal scenarios. 
We also want to further investigate how this approach could be integrated in other efficient public transit routing algorithms.

\bibliographystyle{plain}
\bibliography{arxiv}

\end{document}